# Insulating state in tetralayers reveals an even-odd interaction effect in multilayer graphene

Anya L. Grushina[1], Dong-Keun Ki[1], Mikito Koshino[2], Aurelien A. L. Nicolet[3], Clément Faugeras[3], Edward McCann[4], Marek Potemski[3] and Alberto F. Morpurgo[1*]

[1]*Départment de Physique de la Matiére Condensée (DPMC) and Group of Applied Physics (GAP), University of Geneva, 24 Quai Ernest-Ansermet, CH1211 Genéve 4, Switzerland.*

[2]*Department of Physics, Tohoku University, Sendai, 980–8578, Japan.*

[3]*Laboratoire National des Champs Magnétiques Intenses, CNRS-UJF-UPS-INSA, Grenoble 38042, France*

[4]*Department of Physics, Lancaster University, Lancaster, LA1 4YB, UK*

[*]e-mail: Alberto.Morpurgo@unige.ch

**The absence of an energy gap separating valence and conduction bands makes the low-energy electronic properties of graphene and its multi-layers sensitive to electron-electron interactions. In bilayers, for instance, interactions are predicted to open a gap at charge neutrality, turning the system into an insulator, as observed experimentally. In mono and (Bernal-stacked) trilayers, interactions, although still important, do not have an equally drastic effect, and these systems remain conducting at low temperature. It may be expected that interaction effects become weaker for thicker multilayers, whose behavior should eventually converge to that of graphite. Here we show that this expectation does not correspond to reality by investigating the case of Bernal-stacked tetralayer graphene (4LG). We reveal the occurrence of a robust insulating state in a narrow range of carrier densities around charge neutrality, incompatible with the behavior expected from the single-particle band structure. The phenomenology resembles that observed in bilayers, but the stronger conductance suppression makes the insulating state in 4LG visible at higher temperature. To account for our findings, we suggest a natural generalization of the interaction-driven, symmetry-broken states proposed for bilayers. This generalization also explains the systematic even-odd effect of interactions in Bernal-stacked layers of different thickness that is emerging from experiments, and has implications for the multilayer-to-graphite crossover.**

Close to charge neutrality and at zero magnetic field, $B = 0$, electron-electron interactions in graphene and its multilayers have a strong effect in a very narrow range of electron density[1]. In monolayer graphene, Coulomb interactions renormalize the Fermi velocity of the Dirac fermions, which –in agreement with theoretical predictions[1]– increases as the Fermi level approaches the charge neutrality point (CNP)[2]. In Bernal bilayers, interactions are considered to be responsible for the opening of a gap[1,3-7], and they turn the system into an insulator at low-temperature[8-11]. Thicker Bernal-stacked multilayers have received less attention, and it is only known that Bernal trilayers remain conducting at low temperature (rhombohedral trilayers –on the contrary– have again been found to open a large gap due to interactions)[12]. In all cases –i.e., for mono and bilayers– the effects of interactions are visible experimentally only at charge densities below $2\text{-}3\times10^{10}$ cm$^{-2}$, and their observation requires the study of suspended devices of the highest quality[13,14], since otherwise the magnitude of charge inhomogeneity is too large. By employing a recently developed fabrication process (see Ref. 15 and Methods for details), here we realize very high-quality multi-terminal suspended devices based on Bernal-stacked 4LG, and use them to bring to light a very pronounced insulating state at the CNP due to electron-electron interactions.

Before discussing the detailed characterization of the thickness and the stacking of the layers in our devices, we present the experimental indications of the high device quality and of the unusual highly resistive state at the charge neutrality point (CNP), which can be appreciated already in the most basic transport measurements. Fig. 1a shows the gate voltage ($V_G$) dependence of the two-terminal resistance ($R_{2\text{-}4}$) of two different devices measured at 4.2 K. In both cases, a very high and narrow peak is observed around the CNP, approximately 50 times higher than that observed in suspended monolayer and Bernal-stacked bi/trilayer graphene at the same temperature[2,8,9,12-14] (~0.3 MΩ as compared to 4-8 kΩ; see Fig. S1a in the Supplementary Information (SI); the two devices exhibit identical behavior and in the rest of the manuscript we show the data from one of them). The peak width is extracted from the log-log plot of the conductance $G_{2\text{-}4} = 1/R_{2\text{-}4}$ versus carrier density $n$, as illustrated in Fig. 1b ($n$ is calculated from the applied $V_G$, with the gate capacitance obtained from the analysis of the quantum Hall effect; see discussion of Fig. 2d). The range in which the conductance stays constant upon increasing $n$ –a measure of the carrier density fluctuations[13,14]– extends only up to $n^* \sim 2\text{-}3\times10^9$ cm$^{-2}$, comparable to the best values reported in the literature for suspended graphene of any thickness[2,8-16]. Finally, the multi-terminal configuration allows magneto-transport to be measured in both longitudinal and transverse configurations[15,16]. Already at moderately low $B$, a fully developed integer quantum Hall effect (IQHE) is observed (see Fig. 1c), with integer conductance plateaus (in units of $e^2/h$) in the transverse resistance ($R_{xy}$) and concomitant vanishing of the longitudinal resistance ($R_{xx}$).

The thickness and stacking order of our samples are determined by magneto-Raman spectroscopy. The 2D Raman peak, occurring around 2710 cm$^{-1}$ at $B=0$ when measured with an excitation at $\lambda =$ 514 nm, exhibits the contribution of two components (see inset of Fig. 2a), as expected for 4LG with Bernal stacking[17-19]. This observation, however, is not sufficient to exclude tri and pentalayer graphene unambiguously. To gain conclusive evidence, Raman spectra were collected in the regime of Landau quantization, by applying a finite $B$ to split the single-particle energy spectrum into discrete Landau levels (LLs). In this regime, the G Raman peak exhibits characteristic anti-crossings upon increasing $B$ –an oscillatory shift of the peak position with a concomitant broadening (see Figs. 2a-b)– when the energy of the $E_{2g}$ phonon matches that of specific inter-LL transitions (see Ref. 20 for details). In few layer graphene, whose electronic spectrum consists of several bands[21-26], the effect of coupling is enhanced in the presence of nearly degenerate transitions (i.e., when inter-LL transitions from different bands occur at nearly the same energy)[27], and become measurable even at room temperature (for non-degenerate transitions, the effect of coupling is too weak to be detected). In the simplest approximation –that we will refer to as the effective bilayer model– the single-particle band structure of 4LG consists of two independent bilayer-like bands with different effective masses[21-26], and predicts pairs of nearly degenerate inter-LL transitions to match the phonon energy at $B \sim 5.8$ T, 7.8 T, and 12 T as marked by the blue circles between Figs. 2a-b[27]. As indicated by the blue dashed lines in Figs. 2a-b, the observed features are in virtually perfect agreement with these predictions, whereas they do not match the expectations for tri and pentalayer graphene (respectively, green and red circles between Figs. 2a-b). Magneto-Raman spectroscopy, therefore, unambiguously identifies our devices as made of Bernal stacked 4LG. Additionally, by showing the persistence of Landau quantization at room temperature, these observation provide a first indication of the high quality of our devices.

To characterize the device quality in more detail, and to gain additional understanding of the low-energy electronic properties, we have investigated transport in the quantum Hall regime by taking advantage of the multi-terminal geometry that enables both $R_{xx}$ and $R_{xy}$ to be measured[15,16]. Fig. 2c shows a color plot of $R_{xx}(V_G, B)$, exhibiting clear features originating from the QHE, with Shubnikov-de Haas (SdH) oscillations becoming visible from $B \sim 0.1$T. Through the condition for their visibility ($\mu B >> 1$), we estimate the carrier mobility $\mu$ to be larger than 100,000 cm$^2$/Vs[13,14]. The oscillations evolve into fully developed IQHE states –with plateaus in $R_{xy}$ and vanishing $R_{xx}$– starting from $B \sim 0.3$ T at filling factor $\nu \equiv nh/eB = 8$ (see the white dashed line in Fig. 2c), and subsequently at $\nu = 12$ and 16 (starting from $B \sim 0.4$-$0.5$ T). These states are clearly apparent in Fig. 2d, which shows the longitudinal ($\sigma_{xx}$) and transverse ($\sigma_{xy}$) conductivity, measured as a function of $V_G$ for $B$ in the range 0.475-0.8 T, and plotted as a function of $\nu$ (the gate capacitance is determined by enforcing the scaling of all curves; the good scaling also allows us to reduce noise by averaging[16] as shown in Fig. 2e). These fully developed states at low $B$ follow the sequence expected from the simplest description of the electronic structure of 4LG in terms of two decoupled bilayer-like bands, for which a 16-fold degenerate LL at zero energy needs to be completely filled before electrons occupy (four-fold degenerate) LLs at higher energy[25]. The appearance of the $\nu = 4$ state, which starts from $B=0.8$ T, indicates that at higher fields, the 16-fold degeneracy of the zero energy LLs starts to break. Possible mechanisms responsible for breaking this degeneracy are the effect of next-nearest-layer hopping[28] (that is not considered in the effective bilayer model) or electron-electron interactions[29], for which an analysis of the behavior at larger $B$ provides additional evidence (Fig. 4). Irrespective of these details, the observation of the 16-fold degenerate zero-energy LL provides useful information about the low-energy electronic properties of suspended 4LG.

With the device structure and quality characterized well, we look at the high resistance peak found close to the CNP (Fig. 1a). At the base temperature ($T$ = 250 mK; see Methods), the conductance $G$ becomes unmeasurably small (the resistance is larger than 100 MΩ), as shown in Fig. 3a. The different curves in the figure correspond to two-terminal measurements done using different pairs of contacts (as indicated in the legend; for these high resistance values, four-terminal measurements cannot be done reliably), and show that the strong suppression always occurs in a same range of carrier density $|n| < $ 2-3×10$^{10}$ cm$^{-2}$, irrespective of the measurement configuration. This finding is an indication of the very high device homogeneity, which can be checked directly when working with multi-terminal devices[15,16]. Fig. 3b shows the temperature dependence of $G(V_G)$ measured in one of the two-terminal configurations (other configurations give identical results), showing a pronounced insulating behavior upon lowering $T$. Above 1 K, the minimum conductance is thermally activated $G_{min} = exp(-E_A/2k_BT)$ with an activation energy $E_A$ ~ 15 K (see Fig. 3c), indicating the presence of an energy gap at the CNP. The gap closes very rapidly when the device is gate-biased away from the CNP, as it can be seen in the measurement of the differential conductance ($dI/dV$) as a function of different $V_G$ and applied bias voltage $V_B$ (Fig. 3d). Fig. 3e further shows that at low bias, the $dI/dV$ vanishes, within the accuracy of the measurements, for all bias voltages $V_B$ below a threshold (~1.5 mV, corresponding approximately to the activation energy extracted from the $T$-dependent measurements), above which it increases sharply. In the case of bilayer graphene (BLG), a qualitatively similar behavior has been observed[8,9] (see also section I in the SI) and interpreted in spectroscopic terms, with the applied bias taken to be a direct measure of the energy of the injected electrons[8,9]. It cannot be excluded, however, that the dominant effect of the applied bias is the electrostatic accumulation of charges on the suspended 4LG flake. Indeed, since the gap closes very rapidly upon increasing the charge density away from the CNP, the accumulation of even a very small amount of charges can result in a drastic suppression of the gap, leading to a large and sharp conductance increase.

The occurrence of an insulating state at the CNP is unexpected. It cannot be accounted for in terms of the gapless single-particle band structure of 4LG[21-26], and its explanation calls for the effects of electron-electron interactions (a scenario based on the presence of macroscopic defects such as AB-BA stacking faults –which has been proposed for bilayers[30,31] to explain why some of high quality suspended devices remain conducting at low $T$[32]– is incompatible with our data, and with different observations reported in the literature, see section III in the SI). Indeed, the appearance of fully developed QHE at $B$ = 2.3 T with plateaus in $\sigma_{xy}$ at 1, 2, and 3 $e^2/h$ and vanishing $\sigma_{xx}$ (Fig. 4a) demonstrates that interactions are strong enough to completely lift the degeneracy of the zero-energy LL already at low $B$. Plotting $dG/dB$ as a function of $B$ and $n$ shows that these symmetry broken states start developing from $B$ ~ 1 T (Fig. 4b). This value corresponds to the magnetic field at which the $\nu$ =4 state becomes visible, suggesting that the appearance of this state is also due to interactions. Note that, in the fractional quantum Hall effect (likely to be within experimental reach at higher $B$ with devices of the quality shown here), the 16-fold degeneracy of the zero-energy Landau level is expected to lead to new unexplored regimes, occurring when states belonging to different LLs are mixed by interactions[33-35]. The first manifestations of these new regimes –the appearance of an even-denominator fractional state at $\nu$ = -1/2[16] and of electron-hole asymmetry[36,37]– have just been reported in BLG: with its larger degeneracy, 4LG should lead to an even richer spectrum of new phenomena. Finally, additional evidence for the relevance of the interactions in 4LG is provided by the enhancement of the insulating state around charge neutrality, which at higher magnetic field extends throughout a larger density range (Fig. 4c). This behavior is identical to that observed in mono and bilayer graphene, where it has been shown to originate from a canted anti-ferromagnetic state due to interactions[38-40] (a similar explanation may be valid for 4LG: indeed Fig. 4d, which shows that the insulating state is not affected by a parallel magnetic field $B$ up to 15 T, indicates that the electron spin is not a good quantum number in the insulating state). Finding that the enhancement starts already at low $B$ (Fig. 4c), and that the insulating state at finite field evolves continuously into the $B$=0 insulating state upon reducing $B$, suggests a related origin of both states.

A comprehensive analysis of the effect of electron-electron interactions in Bernal-stacked 4LG is complex, and goes beyond the scope of this paper. However, our observation that the magneto-Raman measurements and the IQHE at low $B$ are in surprisingly good agreement with the behavior expected from the effective bilayer model suggests that treating 4LG by analogy with the case of graphene bilayers is a good starting point. At a mean-field level[3,5,7,8], the low-energy properties of the symmetry-broken states in BLG are commonly described within a two-band model, in terms of a 2×2 matrix Hamiltonian of the type

$$H^{BLG} = -\tfrac{\hbar^2}{2m^*}\left((k_x^2 - k_y^2)\sigma_x \pm 2k_x k_y \sigma_y\right) - \Delta \sigma_z \qquad (1)$$

for each valley and spin (the " ± " sign changes going from the K to the K' valley). The first term is the kinetic energy of the electrons ($m^*$ is the effective mass) and the second one represents the order parameter of the broken symmetry state ($\sigma_{x,y,z}$ are Pauli matrices acting on the layer degree of freedom). Different ground states emerge, depending on how the sign of the order parameter $\Delta$ changes upon switching valley or spin[7,8], but, in all cases, the "bulk" of BLG is gapped (see Fig. 5a) as the term $\Delta\sigma_z$ in the Hamiltonian has an opposite sign in opposite layers for any fixed spin and valley configuration. The analogy with BLG suggests that this mean-field treatment can be generalized to the case of 4LG in a natural way, by describing the low-energy electronic properties in terms of a 4×4 matrix Hamiltonian of the type

$$H_{i,j}^{4LG} = K_{i,j} + (-1)^i \Delta \delta_{i,j} \qquad (2)$$

where the indexes $i, j$ identify the layer ($i, j = 1,\ldots,4$; see Ref. 26 and section II in the SI). The first term is the kinetic energy of electrons and the second one is the order parameter originating from interactions. This extension of the BLG description to 4LG amounts to generalizing the mean field term $\Delta\sigma_z$ to the staggered interlayer potential $(-1)^i\Delta\delta_{i,j}$, which changes its sign between neighboring layers, just like in BLG. It is straightforward to show by direct diagonalization of $H_{i,j}^{4LG}$ that, within the effective bilayer model, the staggered potential $\Delta$ opens a gap at the CNP in 4LG (see Fig. 5a; again in analogy to the BLG, ground states with different properties originate from how $\Delta$ changes upon switching valley or spin). While the above discussion based on the analogy with BLG certainly cannot substitute a complete theoretical analysis, it has the benefit of showing explicitly a realistic scenario based on electron-electron interactions, which leads to the opening of a gap at the CNP and naturally explains why the gap size in 4LG is comparable to that observed in BLG (since within the current approximation, the electronic structure of 4LG is equivalent to that of two decoupled BLG, even in the presence of a staggered potential; see the SI for more details).

Interestingly, the scenario that we propose also captures the systematic behavior of charge neutral Bernal multilayers, which emerges experimentally with our finding of an interaction-driven insulating state in 4LG graphene. Specifically, as it has been reported in previous experiments, "odd" (i.e., mono and tri) layers remain conducting at low temperature[2,12-14], whereas, as we can conclude from this work, "even" (i.e., bi[8-11] and tetra) layers exhibit a gap and become insulating. This phenomenon cannot simply be explained in terms of an enhanced susceptibility to interactions due to a larger low-energy density of states, since Bernal-stacked trilayers, for instance, have a larger density of states than bilayers, and yet they remain conducting at low temperature. On the contrary, describing the effect of interactions in terms of a staggered mean-field potential at the level of approximation of Eqs. (1) and (2) leads precisely to the even-odd effect that is observed experimentally. In fact, at the same level of approximation used to discuss 4LG, the electronic structure of even multilayers generally decouples into bilayer-like bands[21-26], and since each band opens a gap in the presence of a staggered potential, the electronic state of a generic even multilayer is gapped (see Fig. 5a). When performing the same analysis for odd multilayers, next to the bilayer-like bands, a linearly dispersing monolayer-like band emerges and remains ungapped in the presence of a staggered potential (see Fig. 5b). As a result, odd multilayers remain metallic without opening a gap (see section II in the SI for more details). It remains to be understood why the simple approximation used in this study to describe Bernal multilayers (which neglects, for instance, next-layer hopping processes, responsible for the $\gamma_2$ and $\gamma_5$ parameters in the tight-binding description of graphite) correctly captures the basic aspects of the electron-electron interactions. It is likely that the robustness of the even-odd effect is rooted in the symmetry of these systems, which is different in the two cases: even multilayers are inversion symmetric whereas odd ones possess reflection symmetry[26] (for thickness larger than the monolayer). It therefore seems reasonable that the appearance of a staggered potential –which breaks inversion and preserves reflection symmetry– can account for a drastic modification of the behavior of even multilayers, and not of that of odd ones.

An implication of the robustness of the even-odd interaction effect, and of the experimental findings reported here, is that electron-electron interactions can be expected to determine also the properties of multilayers that are so thick to be normally considered as bulk graphite. Indeed, finding an insulating state in 4LG already clearly visible well above 4.2 K, i.e. much more pronounced than in BLG (where it becomes visible only below 1 K; see section I in the SI), shows that the role of interactions becomes stronger, and not weaker, upon increasing layer thickness. This finding and the theoretical considerations made above, therefore, indicate that the commonly held notion that bulk graphite can be well understood within a single particle model is unlikely to be correct, and that at sufficiently low-energy the effects of electron-electron interactions become important. These effects can easily be eclipsed in experiments on samples of microscopic size, which unavoidably contain structural defects, but when probing well-characterized, small multilayers –which can be made defect free– they become fully apparent.

**Methods**

Samples in this study were fabricated by using polydimethylglutarimide (PMGI) based lift-off resist (LOR, MicroChem) as a sacrificial layer[41], following the procedure described in detail in Ref. 15. Graphene flakes were mechanically exfoliated from natural graphite onto an approximately 1-µm-thick LOR layer covering the highly doped Si/SiO$_2$ substrate (acting as a back gate), and the metallic electrodes (10 nm Ti/70 nm Au) contacting the selected flakes were defined by using a standard micro-fabrication process: electron-beam lithography (EBL), deposition, and lift-off. Before removing the sacrificial LOR resist locally under the flake to achieve suspension, the flake shape was defined by exposure to an Oxygen plasma through a patterned polymethylmethacrylate (PMMA) resist mask (see the inset of Fig. 1a for an optical image of etching pattern). After suspension, the devices were mounted in the vacuum chamber of a Heliox $^3$He system (Oxford Instruments; base temperature of 250 mK), and current annealed at low temperature ($T$ = 4.2 K) in a two-terminal configuration, by using shorted adjacent pair of contacts (1 and 2) and (3 and 4) as a source and drain, respectively (see the detailed description of the process in Ref. 15).

**Supplementary Information** accompanies the paper on **www.nature.com/nature.**

**Acknowledgements**

A.L.G., D.K.K. and A.F.M. thank Nicolas Ubrig for help with measuring Raman spectra of the devices at zero magnetic field, and A. Ferreira for technical support. A.F.M. gratefully acknowledges support from the SNF and the NCCR QSIT. M.K. is funded by JSPS Grant-in-Aid for Scientific Research Nos. 24740193 and 25107005. The work has been also supported by the EU under the Graphene Flagship (A.F.M., C.F. and E.M.) and by the European Research Council (A.A.L.N. and M.P.).

**Author contributions**

A.L.G. and D.K.K. fabricated the devices, performed the electrical transport measurements and analysed the data. C.F., A.N., A.L.G., and M.P. carried out the magneto-Raman measurements and analysed the results. M.K. and E.M. contributed to the theoretical analysis, and M.K. performed theoretical calculations of the effect of staggered interlayer potential in multilayer graphene. A.F.M. supervised the work, discussed the transport experiments with A.L.G. and D.K.K. and interpreted the results in collaborations with A.L.G., D.K.K., M.K. and E.M. The paper and the supplementary information were written by A.F.M., A.L.G., D.K.K. and M.K, who included suggestions and comments from all authors.

**Competing financial interests** The authors declare no competing financial interests.

Correspondence and requests for materials should be addressed to A.F.M. ([Alberto.Morpurgo@unige.ch](Alberto.Morpurgo@unige.ch)).

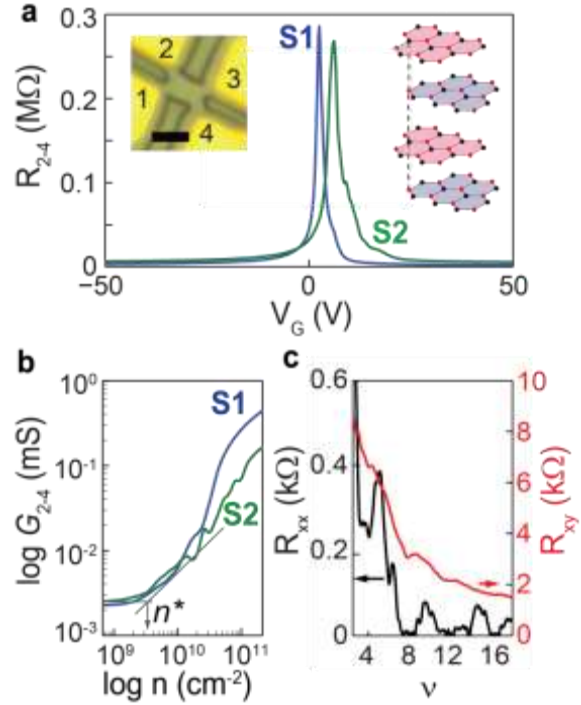

**Fig. 1 | Very high resistance at charge neutrality in high-quality suspended tetralayer graphene devices**. **a**, $V_G$-dependence of the two-terminal resistance $R_{2-4}$ measured in two devices (S1 and S2) at 4.2 K, showing a peak resistance ~0.3 MΩ at the CNP, much larger than the value normally observed in mono or bilayer graphene at the same temperature ($R_{\alpha-\beta}$ represents the two-terminal resistance measured between contacts $\alpha$ and $\beta$, with the indexes labelling the contacts shown in the left inset). Left inset: optical microscope image of a device covered by a PMMA mask prior to etching (see Methods; the scale bar is 2 μm long). Right inset: crystal structure of Bernal-stacked 4LG. **b**, Conductance $G_{2-4} = 1/R_{2-4}$ as a function of $n$ in a double-logarithmic scale, showing a very low level of charge inhomogeneity, $n^* = 2\text{-}3\times10^9$ cm$^{-2}$ (pointed to by the arrow). The density is given by $n = \alpha(V_G - V_{CNP})$, with $\alpha = 4.66\times10^9$ cm$^{-2}$/V determined from the analysis of QHE (see Fig. 2d). **c**, Fully developed QHE measured at $B = 0.45$ T, with vanishing longitudinal resistance $R_{xx} = R_{1-4,2-3}$ (black curve) and quantized transverse resistance $R_{xy} = R_{1-3,4-2} = 1/\nu \times h/e^2$ (red curve) at $\nu = 8$, 12, and 16 (data taken at 250 mK; $R_{\alpha-\beta,\gamma-\delta}$ corresponds to the ratio of the voltage difference measured between contacts $\gamma$ and $\delta$, and the current flowing from contact $\alpha$ to $\beta$).

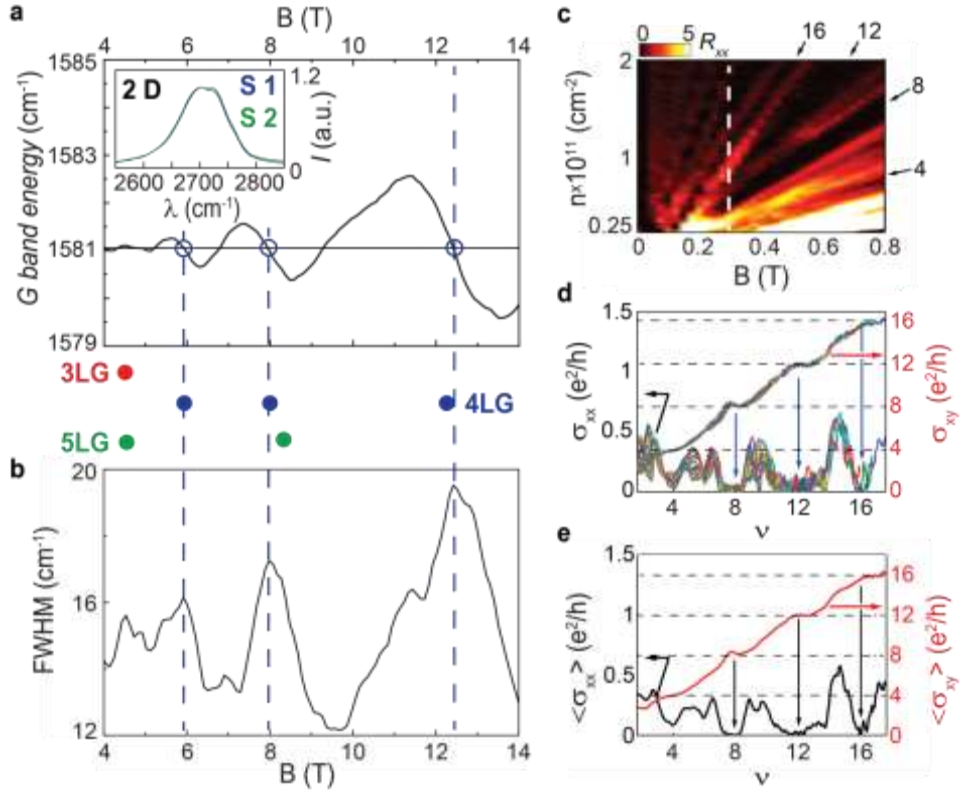

**Fig. 2 | Magneto-Raman spectroscopy and four-probe quantum Hall transport in suspended tetralayer graphene**. **a**, Position of the Raman G-band peak as a function of $B$ measured at room temperature on one of our suspended devices, exhibiting characteristic magneto-Raman oscillations (the inset shows the 2D Raman peak at zero $B$). The empty circles denote the position at which the anti-crossings occur, which coincide at values of $B$ for which the broadening of the Raman G line (measured by the full width at half maximum –FWHM– shown in **b**) peaks. The red, blue and green circles between panels **a** and **b** correspond to the values of $B$ at which the anti-crossings are expected for tri, tetra, and pentalayer Bernal-stacked graphene respectively (see the main text for details): this comparison unambiguously identifies our layer as Bernal-stacked 4LG. **c**, $R_{xx}$ plotted as a function of $B$ and $n$. Clear Shubnikov-de Haas oscillations are visible starting from $B = 0.1$ T, and fully developed quantum Hall states from $B = 0.3$ T. The arrows indicate the local $R_{xx}$ minima at $\nu = 4, 8, 12,$ and $16$ (the $\nu = 8$ state is the first to develop fully at $B \sim 0.3$ T; white dashed line). At low $n$, below $2.5 \times 10^{10}$ cm$^{-2}$, the device becomes highly insulating (see Fig. 3), preventing multi-terminal measurements. **d**, Traces of the longitudinal and transverse conductivities, $\sigma_{xx} = \rho_{xx}/(\rho_{xx}^2 + \rho_{xy}^2)$ and $\sigma_{xy} = \rho_{xy}/(\rho_{xx}^2 + \rho_{xy}^2)$, as a function of $\nu$, measured for $B$ between 0.475 T and 0.8 T ($\rho_{xx} = R_{xx} \times W/L$ and $\rho_{xy} = R_{xy}$ with the device aspect ratio $W/L = 0.87$). The filling factor $\nu = nh/eB$ is calculated from $B$ and $n = \alpha(V_G - V_{CNP})$ with $\alpha = 4.66 \times 10^9$ cm$^{-2}$/V determined by imposing scaling of all the measured curves in this magnetic field-range. The good scaling enables averaging of the magneto-conductivity curves to suppress noise, as shown in **e**.

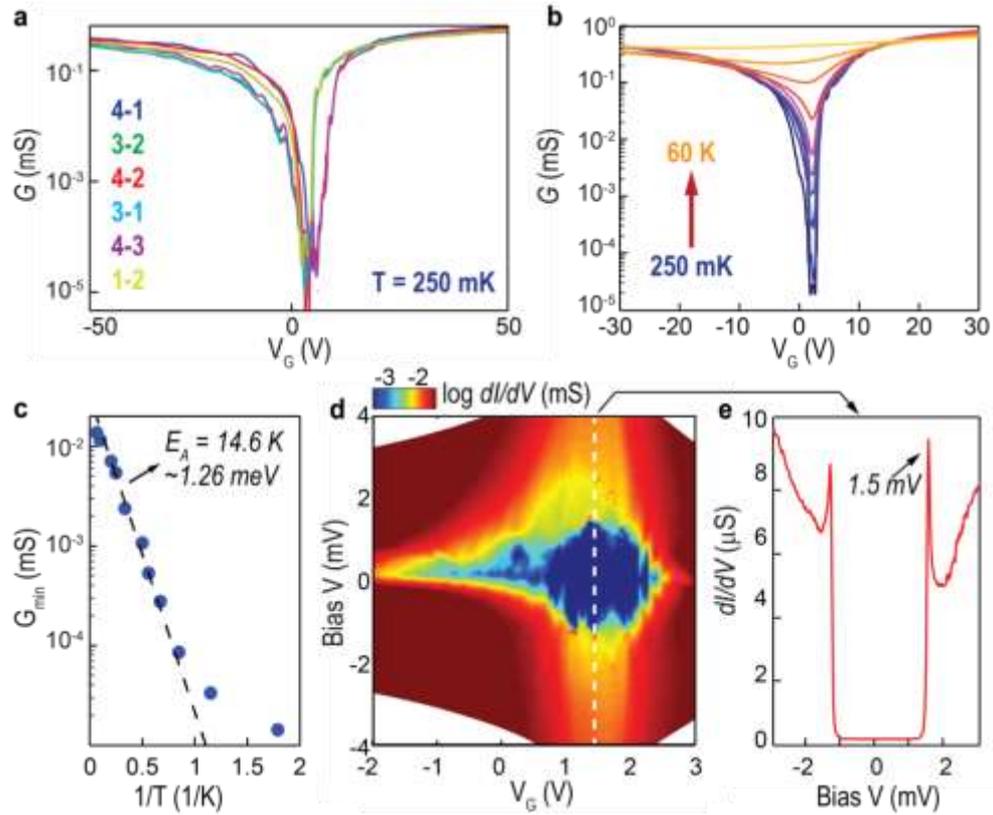

**Fig. 3 | Insulating state at charge neutrality and $B = 0$ T in tetralayer graphene. a**, $\log(G)$ versus $V_G$, measured in several different two-terminal configurations, as indicated in the legend (see inset in Fig. 1a for the contact labelling). The same behavior –a pronounced suppression of $G$ in a very narrow range of gate voltages $|V_G - V_{CNP}| < 5$ V, corresponding to $|n| < 2\times10^{10}$ cm$^{-2}$– is observed irrespective of the contacts used, which is indicative of the high device homogeneity. **b**, $T$-dependence of $\log(G)$-versus-$V_G$ measured in a two terminal configuration. Upon increasing $T$, the minimum conductance $G_{min}$ at the CNP increases by more than four orders of magnitude, exhibiting a thermally activated behavior for $T > 1$ K, as shown in panel **c** ($G_{min} = exp(-E_A/2k_BT)$ with $E_A \sim 14.6$ K). **d**, Color plot of the differential conductance $G = dI/dV$ (in log scale) as a function of source-drain ($V_B$) and gate voltage ($V_G$), measured at 250 mK. The insulating state corresponds to the dark-blue region close to the CNP and $V_B = 0$ V. **e**, Line-cut, $dI/dV(V_B)$, of the color plot shown in **d**, taken at $V_G = 1.6$ V (corresponding to the white dashed line in panel **d**). Below $V_B \sim 1.5$ mV (indicated by the arrow), the conductance vanishes within the accuracy of the measurement.

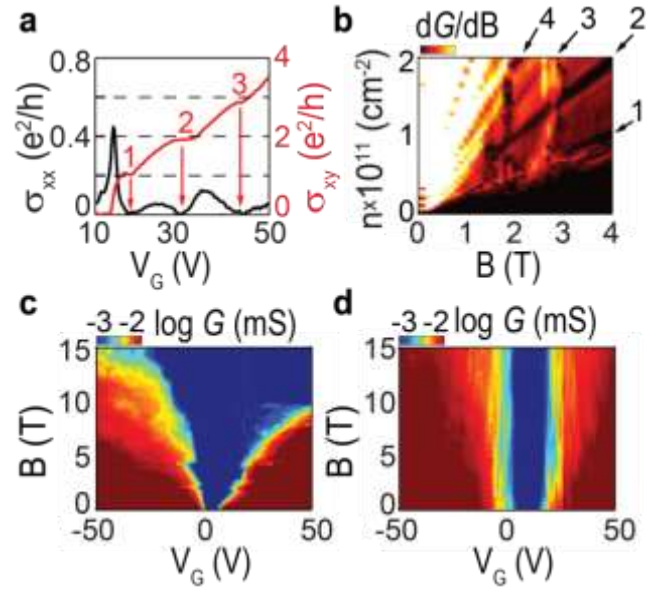

**Fig. 4 | Strong electron-electron interactions at finite magnetic field. a**, $V_G$-dependence of the longitudinal ($\sigma_{xx}$) and the transverse ($\sigma_{xy}$) conductivity at $B = 2.3$ T, showing fully developed broken symmetry states at $\nu = 1$, 2, and 3 with zeros in $\sigma_{xx}$ and plateaus in $\sigma_{xy}$. **b**, Fan-diagram of the derivative of the conductance $dG/dB(B, n)$, illustrating the evolution of the quantum Hall states at $\nu = 1$, 2, 3, and 4 (indicated by arrows) as a function of $B$ and $n$: these states survive down to ~1 T (the vertical features visible at $B$ around 1.8 and 2.8 T are likely due to crossing in energy between the different symmetry broken Landau levels). **c**, Upon the application of perpendicular magnetic field, the insulating state extends through a much broader range of carrier density, similarly to what happens in mono and bilayer graphene. **d**, If the magnetic field is applied in the plane, no change is seen in the range of $B$ accessible in our laboratory (15 T). All measurements shown in this figure were taken at $T = 250$ mK.

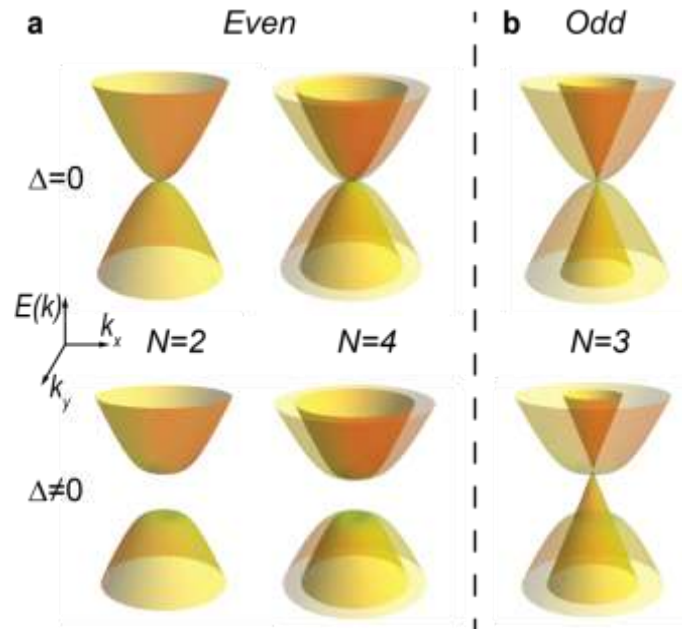

**Fig. 5 | Even-odd effect of the staggered interlayer potential Δ on the band structures of few layer graphene. a**, Band structure of few layer graphene with an even number of layers ($N = 2$ on the left and $N = 4$ on the right) in the absence (upper panels) and presence (lower panels) of a staggered interlayer potential Δ (calculations are based on a tight-binding approximation, including only nearest neighbour hopping in the plane and perpendicular to the plane). The opening of a gap at the charge neutrality point at finite Δ is apparent. On the contrary, no gap is opened by a staggered potential Δ in the case of few layer graphene with odd $N$, as shown in **b** for the case of trilayer graphene ($N = 3$), because of the presence of a monolayer-like band (see section II in the SI for the theoretical calculations and more details).